\documentclass[%
 reprint,
 superscriptaddress,
 amsmath,amssymb,
 aps,
 prb,
 floatfix
]{revtex4-2}

\usepackage{booktabs}
\usepackage{graphicx}
\usepackage[colorlinks=true,linkcolor=magenta,citecolor=magenta]{hyperref}
\usepackage{microtype}
\usepackage{multirow}
\usepackage{xcolor}
\usepackage[normalem]{ulem}

\newcommand\dee{d^{\vphantom\dagger}}
\newcommand\ddg{d^\dagger}
\newcommand\cee{c^{\vphantom\dagger}}
\newcommand\cdag{c^\dagger}

\newcommand\ii{\mathrm{i}}
\newcommand\ee{\mathrm{e}}
\newcommand\iv{\ii\nu}
\newcommand\ivv{\ii\vec\nu}
\newcommand\iw{\ii\omega}
\newcommand\eye{\mathbf{1}}
\newcommand\TT{\mathcal{T}}

\newcommand\emax{\epsilon_{\mathrm{max}}}
\newcommand\bigO{\mathcal{O}}
\DeclareMathOperator{\sgn}{sgn}
\DeclareMathOperator{\tr}{tr}
\newcommand\primedsum{\sideset{}{'}\sum}

\newcommand\AddAuthorComment[3]{%
    {\color{#1} ({\bf #2}%
        \if\relax\detokenize{#3}\relax%
        \else%
            {\normalfont: #3}%
        \fi%
    )}%
}
\newcommand\AuthorReplace[5]{%
    \AddAuthorComment{#1}{#2}{#3}%
    \if\relax\detokenize{#4#5}\relax%
    \else%
        { \color{#1}\sout{#4}\uwave{#5}}
    \fi%
}

\renewcommand\vec{\boldsymbol}

\graphicspath{%
    {figures/}%
    }

\begin{document}

\title{
Overcomplete intermediate representation of two-particle Green's functions and its relation to partial spectral functions
}

\author{Selina~Dirnb\"ock}
\affiliation{Department of Solid State Physics, TU Wien, 1040 Vienna, Austria}

\author{Seung-Sup B.~Lee}
\affiliation{Department of Physics and Astronomy and Center for Theoretical Physics, Seoul National University, Seoul 08826, Korea}

\author{Fabian~B.~Kugler}
\affiliation{Center for Computational Quantum Physics, Flatiron Institute, 162 5th Avenue, New York, NY 10010, USA}

\author{Sebastian~Huber}
\affiliation{Department of Solid State Physics, TU Wien, 1040 Vienna, Austria}

\author{Jan~von~Delft}
\affiliation{Arnold Sommerfeld Center for Theoretical Physics, 
Center for NanoScience,\looseness=-1\,  and 
Munich Center for \\ Quantum Science and Technology,
Ludwig-Maximilians-Universit\"at M\"unchen, 80333 Munich, Germany}

\author{Karsten~Held}
\affiliation{Department of Solid State Physics, TU Wien, 1040 Vienna, Austria}

\author{Markus Wallerberger}
\affiliation{Department of Solid State Physics, TU Wien, 1040 Vienna, Austria}

\date{\today}

\begin{abstract}
Two-particle response functions are a centerpiece of both experimental and theoretical quantum many-body physics. Yet, due to their size and discontinuity structure, they are challenging to handle numerically. Recently, two advances were made to tackle this problem: first, the overcomplete intermediate representation (OIR), which provides a highly efficient compression of Green's functions in imaginary frequency, and second, partial spectral functions (PSFs), which allow for an efficient evaluation in real frequency.
We show that there is a two-to-one correspondence between PSFs and OIR coefficients and exploit this fact to construct the OIR for three-or-more-particle propagators. We then use OIR to fit and compress imaginary-frequency data obtained from the numerical renormalization group (NRG), reaching a compression ratio of more than 400. Finally, we attempt to match the OIR data to partial Green's functions from NRG.Due to the overcompleteness, we achieve only qualitative agreement. 
\end{abstract}

\maketitle

\section{Introduction}
\label{sec:intro}

Green's functions (GFs) are a critical tool for understanding the physics of quantum many-body systems. One-particle GFs relate to spectral functions, amendable to spectroscopy experiments, while higher-order, two-or-more-particle GFs relate to linear and nonlinear response functions.  Naturally, they also form the basis of a smorgasbord of many-body frameworks~\cite{Economou2006}. While in analytic calculations, we frequently mix GFs of all orders, higher-order GFs are considerably more intricate when working with them numerically. This comes down to two problems: one of space and one of structure.

The first problem, space, is simply the curse of dimensionality: the memory required to naively store the simultaneous movement of $n$ quantum particles scales exponentially in $n$. Current solutions have focused on making the base of that exponent as small as possible: when working in imaginary time, one can construct an almost maximally compact basis, the so-called intermediate representation (IR)~\cite{Shinaoka17:compressing,Li20}. For the dependence on position and (real) time, quantics tensor trains provide a controlled and, at least in some cases, very compact representation~\cite{ShinaokaPRX23,Ritter23}. These and other~\cite{FernandezPRX22} tensor trains can also be used in an attempt to cure the exponential scaling itself.

The second problem, the structure of many-body GFs, is linked to the quantum nature of the underlying particles: the (anti-)commutativity of bosons (fermions) causes discontinuities at equal-time planes. For higher-order GFs, some of these planes run ``diagonally'' through the time domain~\cite{Wallerberger2016}, which implies that any discretization given by
a direct product of the single-particle basis cannot be compact. One can mitigate this by subtracting the jumps, either numerically~\cite{GangLi16} or diagrammatically~\cite{Wentzell20,Krien2021}, but this still leaves non-analyticities in these locations. Alternatively, one may elect to not store these GFs at all, but compute them on-the fly, either stochastically~\cite{Prokofev08} or analytically~\cite{Taheridehkordi19}.

Two recent approaches address the structure problem directly: in Refs.~\cite{Shinaoka18:overcomplete,Wallerberger21:BSE}, the two-particle imaginary-frequency GFs is represented as a sum of twelve separate terms identified by their analytic form, each of which is smooth. This admits the construction of an almost maximally compact, albeit overcomplete, intermediate representation (OIR).
In Ref.~\cite{Kugler21}, an arbitrary $n$-point ($\lceil n/2\rceil$-particle) GFs, in real or imaginary frequencies, is represented as a sum of $n!$ terms.
Each of these terms, to be called partial Green's functions (PGFs),
is the convolution of a simple, system-independent integral kernel with a partial spectral function (PSF). By computing the PSFs, e.g., via exact diagonalization~\cite{Tanaka19} or the numerical renormalization group (NRG)~\cite{Lee21PRX}, and the resulting PGFs separately rather than as a sum, one again can work with more compact discretizations.

The natural questions arising from this, which we shall address in Sec.~\ref{sec:psf}, are the following: is there a connection between the OIR and PGFs? And, if so, can we use the PGFs, which were derived for all orders \cite{Kugler21,Halbinger23}, to construct the corresponding OIR, which has previously only been done for two-particle quantities?

Having answered these questions in the affirmative, we will move to a more subtle point in Sec.~\ref{sec:results}: the coefficients of the OIR are usually fitted while the PGFs are computed. Since the OIR is overcomplete, there is an ambiguity in exactly how we determine the coefficients, in other words, the corresponding fitting problem is poorly conditioned. This is no problem for the OIR itself. The question is, when fitting the OIR in imaginary frequencies, do its constituents still match the original PGFs? Section~\ref{sec:conclusions} offers our conclusions and an outlook.

\section{Partial Green's functions and overcompleteness}
\label{sec:psf}

In this section, we aim to connect two descriptions of the multipoint GFs: (a)
PSFs and PGFs~\cite{Kugler21}, which originate from considering all possible permutations of operators, and (b) the OIR~\cite{Shinaoka18:overcomplete}, which originates from grouping terms in the Lehmann representation in imaginary frequency by different kernels. 

For completeness, we review two-point imaginary-frequency GFs and the compression of such objects~\cite{Shinaoka17:compressing,Li20} in Secs.~\ref{sec:imag2} and \ref{sec:ir2}. This sets the stage for our two main results: (i) establishing a two-to-one connection between PGFs and the terms in the OIR for the arbitrary $n$-point case in Sec.~\ref{sec:imagn}, and (ii) using this to generalize the overcomplete basis \cite{Shinaoka18:overcomplete} to the general $n$-point case and showing how to obtain the coefficients in Sec.~\ref{sec:irn}.

\subsection{Two-point partial Green's functions}
\label{sec:imag2}
Let us start with the two-point GF in imaginary time for simplicity. (Most of this material is well-known but serves to introduce the topic and our notation.) Its definition is:
\begin{equation}
    \tilde G(\tau_1, \tau_2) := -\sum_\psi \frac{\ee^{-\beta E_\psi}}Z
    \langle \psi |\TT A_1(\tau_1) A_2(\tau_2) |\psi \rangle,
\label{eq:G2tau}
\end{equation}
where $A_i$ are fermionic operators;
$\tau_i$ are imaginary (Euclidean) times, which can be restricted to $0 \le \tau_i \le \beta$, where $\beta^{-1}$ is temperature;
imaginary-time evolution is governed by $A_i(\tau)=\ee^{H\tau} A_i \ee^{-H\tau}$,
where $H$ is the Hamiltonian;
$H|\psi\rangle = E_\psi|\psi\rangle$ defines an eigenstate $\psi$ and its energy $E_\psi$;
$Z:=\tr\exp(-\beta H)$ is the grand canonical partition function,
and the chemical potential was absorbed into the Hamiltonian.

The effect of the time-ordering symbol $\TT$ on the expectation value in 
Eq.~\eqref{eq:G2tau} is to split it up into a sum over two operator permutations:
\begin{equation}
\begin{split}
    \tilde G(\tau_1,\tau_2) =  \sum_\psi \frac{\ee^{-\beta E_\psi}}Z\!%
    \begin{cases}
        -\langle\psi |A_1(\tau_1) A_2(\tau_2) |\psi\rangle, &\tau_1 > \tau_2, \\
        +\langle\psi |A_2(\tau_2) A_1(\tau_1) |\psi\rangle, &\tau_1 < \tau_2.
    \end{cases}
\end{split}
\label{eq:G2tau-expl}
\end{equation}
To condense the equations, we introduce the following notation~\cite{Kugler21}: by $\bar1\bar2\in\{12,21\}$ we denote a permutation of the indices $12$. For the trivial
permutation, e.g., we have $\bar1\bar2=12$, and so replace $\bar1$ with $1$ and $\bar2$ with $2$; for the reversed one, we have $\bar1\bar2=21$ and replace
$\bar 1$ with $2$, and $\bar2$ with $1$. By $\sgn(\bar1\bar2)$ we denote the sign of the permutation, and by $\sum_{\bar1\bar2}$ the sum over all permutations,
$\bar1\bar2\in \{12, 21\}$.
Using this notation, Eq.~\eqref{eq:G2tau-expl} becomes:
\begin{equation}
    \tilde G(\tau_1,\tau_2) = \sum_{\bar1\bar2} \tilde G_{\bar1\bar2}(\tau_{1}, \tau_{2}),
    \label{eq:G2tau-split2}
\end{equation}    
where $\tilde G_{12}$ and $\tilde G_{21}$ are imaginary-time PGFs, defined as:
\begin{equation}
\begin{split}
   \tilde G_{\bar1\bar2}(\tau_1,\tau_2) &:= -\Theta(\tau_{\bar1}-\tau_{\bar2}) \sgn(\bar1\bar2) \\
     &\times \sum_\psi \frac{\ee^{-\beta E_\psi}}Z
    \langle\psi |A_{\bar1}(\tau_{\bar1}) A_{\bar2}(\tau_{\bar2}) |\psi\rangle.
\end{split}
    \label{eq:PC2tau}
\end{equation}

We define the Fourier transform of Eq.~(\ref{eq:G2tau}) as:
\begin{equation}
    G(\iv_1, \iv_2) := \int_0^\beta d^2\tau\ \ee^{\iv_1\tau_1 + \iv_2\tau_2} \tilde G(\tau_1, \tau_2),
    \label{eq:g2iv-def}
\end{equation}
where $\iv_1$ and $\iv_2$ are fermionic imaginary or Matsubara frequencies,
$\iv\in \{\frac{\ii\pi}\beta(2k+1)\}$, and $k$ is some integer. One can perform
the Fourier transform by substituting $(u_1, u_2)\equiv(\tau_{\bar1}-\tau_{\bar2}, \tau_{\bar2})$
into each PGF~(\ref{eq:PC2tau}), which leads to:
\begin{equation}
    G(\iv_1,\iv_2) = \sum_{\bar1\bar2} G_{\bar1\bar2}(\iv_1, \iv_2),
\label{eq:G2iv}
\end{equation}
with the PGFs (\ref{eq:PC2tau}) in Matsubara frequencies reading:
\begin{equation}
\begin{split}
    G_{\bar1\bar2}(\iv_1,\iv_2) &= \beta\delta_{\iv_1+\iv_2,0} \sgn(\bar1\bar2) \\
    &\times \sum_\psi \frac{\ee^{-\beta E_\psi}}Z
    \langle\psi |A_{\bar1} \frac 1{\iv_{\bar1} + E_\psi - H} A_{\bar2} |\psi\rangle.
\end{split}
\label{eq:PC2iv}
\end{equation}
In Eq.~(\ref{eq:PC2iv}), we adopted the common convention of understanding the reciprocal $1/(z-H)$ as the resolvent $(z\eye - H)^{-1}$, where $\eye$ is the identity.

We can now separate the system-dependent part of Eq.~(\ref{eq:PC2iv}) into PSFs $\rho_{12}$
and $\rho_{21}$, defined as~\cite{Kugler21}:
\begin{equation}
    \rho_{\bar1\bar2}(\epsilon) :=
    \sgn(\bar1\bar2) \sum_\psi \frac{\ee^{-\beta E_\psi}}Z
    \langle\psi |A_{\bar1} \delta(\epsilon + E_\psi - H) A_{\bar2} |\psi\rangle,
\label{eq:rho2}
\end{equation}
where $\delta$ is the Dirac delta generalized to operator arguments: $\delta(z - H) = \sum_\phi \delta(z - E_\phi) |\phi\rangle\langle\phi|$.
This permits us to represent the PGFs~(\ref{eq:PC2iv}) as a simple convolution:
\begin{equation}
G_{\bar1\bar2}(\iv_1,\iv_2) = \beta\delta_{\iv_1+\iv_2,0} \int d\epsilon\ 
    \frac{\rho_{\bar1\bar2}(\epsilon)}{\iv_{\bar1} - \epsilon}.
\label{eq:PC2iv-conv}
\end{equation}
From the next section onward, we will assume that the integral can be restricted to some finite interval $[-\emax, \emax]$, or, equivalently, that $H$ is bounded. (This restriction can be relaxed.)

Inserting the PSFs~(\ref{eq:rho2}) into the (full) GF~(\ref{eq:G2iv}) yields
\begin{equation}
    G(\iv_1,\iv_2) = \beta\delta_{\iv_1+\iv_2,0} \int d\epsilon 
     \left[ \frac{\rho_{12}(\epsilon)}{\iv_1 - \epsilon} + \frac{\rho_{21}(\epsilon)}{\iv_2 - \epsilon} \right].
    \label{eq:G2iv-conv}
\end{equation}
Observing $\iv_2 = -\iv_1$ and changing variables $\epsilon \to -\epsilon$ in the second term, we can condense Eq.~(\ref{eq:G2iv-conv}) to a single convolution, the spectral representation of $G$:
\begin{equation}
    G(\iv_1,\iv_2) = \beta\delta_{\iv_1+\iv_2,0} \int d\epsilon \frac{\rho'(\epsilon)}{\iv_1 - \epsilon},
    \label{eq:G2iv-spec}
\end{equation}
where $\rho'$ is the (full) spectral function:
\begin{equation}
    \rho'(\epsilon) = \rho_{12}(\epsilon) - \rho_{21}(-\epsilon).
    \label{eq:rho2prime}
\end{equation}
We shall make note of this fact: the sum of two PSFs forms the argument for a single
convolution in the spectral representation of the imaginary-frequency GF.

For illustration, consider the case that $A_1=\dee$ and $A_2=d^\dagger$ are the annihilation and creation operators for a fermion in some spin-orbital, respectively, and $\beta=\infty$. Then, $\rho_{12}(\epsilon)$ and $-\rho_{21}(\epsilon)$ are nonzero for $\epsilon\geq 0$ only and yield the particle- and hole-side of the spectral function $\rho'(\epsilon)$, respectively.

\subsection{Intermediate representation for two-point Green's functions}
\label{sec:ir2}

The numerical transformation (\ref{eq:G2iv-spec}) between $\rho'(\epsilon)$ and $G(\iv)$ is lossy.
This is evident from the singular value expansion of the corresponding kernel~\cite{Bryan90,Otsuki17}:
\begin{equation}
    \frac 1{\iv - \epsilon} = \sum_{\ell=0}^\infty U_\ell(\iv)\,S_\ell\,V_\ell(\epsilon),
    \label{eq:sve}
\end{equation}
where $\{U_\ell\}$ are the left-singular functions, which form an orthonormal set in imaginary
frequencies, $\{V_\ell\}$ are the right-singular functions, which form an orthonormal set in real frequencies.
$S_\ell$ are the singular values, which for a kernel of finite support $[-\emax,\emax]$
decay faster than exponentially with $\ell$~\cite{Chikano18,SciPost21}, epitomizing the loss
of significance from the real to imaginary frequencies.
$V_\ell$ are bounded in $[-\emax,\emax]$.

This loss of information allows for an extremely compact representation of the
Matsubara GF~\cite{Shinaoka17:compressing}, called the IR.
Inserting Eq.~(\ref{eq:sve}) into Eq.~(\ref{eq:G2iv-spec}) yields:
\begin{equation}
    G(\iv_1,\iv_2) \approx \beta\delta_{\iv_1+\iv_2,0} \sum_{\ell=0}^{L-1} g_\ell U_\ell(\iv_1),
    \label{eq:GIR}
\end{equation}
where $g_\ell = S_\ell \int d\epsilon V_\ell(\epsilon) \rho'(\epsilon)$ is
a basis expansion coefficient.  The number of coefficients needed to represent a given GF with a relative error of at most $\varepsilon$ scales as 
$L\sim \log(\beta\emax)\log(\varepsilon^{-1})$~\cite{Chikano18}.

The expansion coefficients $g_\ell$ can also be inferred from imaginary-frequency data using sparse sampling~\cite{Li20}. The kernel (\ref{eq:sve}) determines a set of $L$ frequencies
$\mathcal V_2 = \{\iv_1, \ldots \iv_L\}$ such that Eq.~(\ref{eq:GIR}) can be turned into a well-conditioned
least-squares problem:
\begin{equation}
    \min_{g_\ell} \sum_{\iv\in\mathcal V_2} \bigg| G(\iv, -\iv) - \sum_{\ell=0}^{L-1} g_\ell U_\ell(\iv) \bigg|^2.
\end{equation}
A similar procedure exists for imaginary-time data.

Once we obtained the GF in the IR, its analytic continuation to real frequencies is trivial: $\rho'(\epsilon) = \sum_{\ell=0}^\infty V_\ell(\epsilon) g_\ell/S_\ell $, though one must regularize this expression heavily due to the rapid decay of the singular values.
Note that, in an ``observed'' imaginary-fre\-quen\-cy GF $G(\iv,-\iv)$, the two PSFs are combined according to Eq.~(\ref{eq:G2iv-spec}). Thus, the IR, which relies on a decomposition of the kernel (\ref{eq:sve}), invariably mixes the PSFs, and only the full GF, rather than the partial ones, are accessible.

\subsection{Partial Green's functions in the \texorpdfstring{$n$}{n}-point case}
\label{sec:imagn}
Let us repeat the calculation in Sec.~\ref{sec:imag2} for the $n$-point GF.
Its definition is:
\begin{equation}
    \tilde G(\vec\tau) := (-1)^{n-1} 
    \sum_\psi \frac{\ee^{-\beta E_\psi}}Z
    \langle \psi |\mathcal T A_1(\tau_1) \cdots A_n(\tau_n) |\psi \rangle,
\label{eq:Gtau}
\end{equation}
where $\vec \tau=(\tau_1,\ldots, \tau_n)$ collects imaginary (Euclidean) times, which
we again restrict to $\tau_i \in [0, \beta]$, and $\mathcal T$ again orders 
operators by imaginary time.

The Fourier transform of Eq.~(\ref{eq:Gtau}) is defined as:
\begin{equation}
    G(\ivv) := \int_0^\beta d^n\tau\,\ee^{\iv_1\tau_1 + \ldots + \iv_n\tau_n} \tilde G(\vec \tau),
\end{equation}
where $\ii\vec \nu=(\iv_1,\ldots, \iv_n)$ collects fermionic Matsubara frequencies.
Using a similar reasoning as in the two-point case (\ref{eq:G2iv}), one finds after a lengthy calculation~\cite{Tanaka19,Kugler21}:
\begin{equation}
\begin{split}
    G(\ivv) 
    &= \sum_{\bar 1\ldots \bar n} G_{\bar 1\ldots \bar n}(\ivv).
\end{split}
\label{eq:Giw-sum}
\end{equation}
Instead of two PGFs as in the two-point case (\ref{eq:PC2iv}), we have $n!$ PGFs:
\begin{equation}
\begin{split}
    & G_{\bar 1\ldots \bar n}(\ivv) = \beta \delta_{\vec\nu} \sgn(\bar1\ldots\bar n)\\
    &\times \sum_\psi \frac{\ee^{-\beta E_\psi}}Z 
     \langle \psi | A_{\bar 1} \prod_{i=1}^{n-1}\!\left[  \frac{1}{\sum_{k=1}^{i} \iv_{\bar k} + E_\psi - H} A_{\overline{i+1}} \right]  | \psi\rangle,
\end{split}
\label{eq:Giw}
\end{equation}
where $\delta_{\vec{\nu}}$ is equal to one if the sum of all frequencies in $\vec{\nu}$ is zero and equal to zero otherwise, $\bar 1\ldots\bar n$ is a permutation, and $\sgn$ denotes its sign.
The resolvent in Eq.~(\ref{eq:Giw}) contains sums of fermionic frequencies; for even $i$,
this gives a bosonic Matsubara frequency, $\iw \in \{\frac{\ii\pi}\beta(2k)\}$, where $k$ is
some integer. As bosonic Matsubara frequencies can be exactly zero, one must either avoid
the poles Eq.~(\ref{eq:Giw}) by carefully taking limits or treat these terms
separately~\cite{Shinaoka18:overcomplete,Kugler21,Halbinger23}.

Instead of just two as in Eq.~(\ref{eq:rho2}), we now have $n!$ PSFs:
\begin{equation}
\begin{split}
    &\rho_{\bar1\ldots\bar n}(\epsilon_1, \ldots, \epsilon_{n-1}) = \sgn(\bar1\ldots\bar n) \\
    &\quad \times\sum_\psi \frac{\ee^{-\beta E_\psi}}Z
    \langle \psi | A_{\bar 1} \prod_{i=1}^{n-1} \!\big[\delta(\epsilon_i + E_\psi - H)\,A_{\overline{i+1}} \big]  | \psi\rangle.
\end{split}
\label{eq:rhon}
\end{equation}

Inserting the PSFs~(\ref{eq:rhon}) into the Fourier transform~(\ref{eq:Giw}), we again find convolutions:
\begin{equation}
\begin{split}
    & G_{\bar 1\ldots \bar n}(\ivv) = \beta\delta_{\vec\nu} \\
    &\quad \times
    \int \frac{ d^{n-1}\vec{\epsilon}\ \rho_{\bar1\ldots\bar n}(\epsilon_1, \ldots, \epsilon_{n-1})}%
    {(\iv_{\bar1} - \epsilon_1)\cdots(\iv_{\bar1} + \ldots + \iv_{\overline{n-1}} - \epsilon_{n-1})}.
\end{split}
\label{eq:Giw-conv-all}
\end{equation}
Here, conservation of energy implies that reversing a permutation $\bar i\to\overline{n + 1 - i}$ together with the order of energies $\epsilon_i \to -\epsilon_{n-i}$ leaves the integral kernel (the denominator of above equation) invariant.  Similarly as in the two-point case where two PSFs can be combined to the full spectral function through Eq.~(\ref{eq:rho2prime}), we can combine pairs of PSFs to what we shall call {\em semipartial spectal functions} (semi-PSFs):
\begin{equation}
\begin{split}
    \rho'_{\bar1\ldots\bar n}(\epsilon_1,\ldots,\epsilon_{n-1}) &:= 
    \rho_{\bar1\ldots\bar n}(\epsilon_1,\ldots,\epsilon_{n-1}) \\ &
    + (-1)^{n-1} \rho_{\bar n\ldots\bar 1}(-\epsilon_{n-1},\ldots,-\epsilon_1).
\end{split}
\label{eq:rhoprime}
\end{equation}
Correspondingly, the {\em semipartial Green's function} (semi-PGF) is defined as:
\begin{equation}
\begin{split}
    & G^\prime_{\bar 1\ldots \bar n}(\ivv) := \beta\delta_{\vec\nu} \\
    &\quad \times
    \int \frac{ d^{n-1}\vec{\epsilon}\ \rho'_{\bar1\ldots\bar n}(\epsilon_1, \ldots, \epsilon_{n-1})}%
    {(\iv_{\bar1} - \epsilon_1)\cdots(\iv_{\bar1} + \ldots + \iv_{\overline{n-1}} - \epsilon_{n-1})},
\end{split}
\label{eq:Giw-conv-prime}
\end{equation}
which equals the full GF for $n = 2$. Otherwise, the (full) GF can be expressed in terms of the semi-PGFs:
\begin{equation}
    G(\ivv) = \primedsum_{\bar 1\ldots \bar n} G^\prime_{\bar 1\ldots \bar n}(\ivv),
    \label{eq:Giw-conv}
\end{equation}
where the primed sum $\sum\!{\vphantom\sum}'$ runs over $n!/2$ inequivalent permutations obtained
under the equivalence $\bar 1\bar 2\ldots \bar n \equiv \bar n\ldots\bar 2\bar 1$.
Table~\ref{table:IRtoPerm_Group} summarizes the corresponding terms in the primed sum for the
case $n=4$.

\begin{table}
\begin{center}
\begin{tabular}{cccc}
 \toprule
  $r$ & \,$\bar1\bar2\bar 3\bar 4$\, & \,$\bar 4\bar 3\bar2\bar 1$\, & coset \\
 \midrule
 1 & 1234 & 4321 &  A \\
 2 & 1243 & 3421 &  B \\
 3 & 1324 & 4231 &  C \\
 4 & 1342 & 2431 &  B \\
 5 & 1423 & 3241 &  C \\
 6 & 1432 & 2341 &  A \\
 7 & 2134 & 4312 &  B \\
 8 & 2143 & 3412 &  A \\
 9 & 2314 & 4132 &  C \\
 10 & 2413 & 3142 & C \\
 11 & 3124 & 4213 & B \\
 12 & 3214 & 4123 & A \\
 \bottomrule
\end{tabular} 
\caption{ Each semi-PGF or semi-PSF with representation index $r$~\cite{Wallerberger21:BSE} (first column) combines the contribution from a pair of permutations, $\bar1\bar2\bar 3\bar 4$ (second column) and its reverse $\bar 4\bar 3\bar2\bar 1$ (third column), from the permutation group $\mathcal S_4$.
The $4!/2 = 12$ pairs are further grouped into three cosets (fourth column).
In each coset, the $\bar1\bar2\bar 3\bar 4$ permutations are related by cyclicity~\cite{Kugler21}; see Sec.~\ref{ssec:cosets} for details.}
\label{table:IRtoPerm_Group}
\end{center} 
\end{table}

\subsection{Overcomplete intermediate representations for the \texorpdfstring{$n$}{n}-point Green's function}
\label{sec:irn}

Equations~(\ref{eq:rhoprime}) to (\ref{eq:Giw-conv}) allow us to construct a compact basis
for an arbitrary $n$-point GF. The crucial observation, made for three- and
four-point functions in Ref.~\cite{Shinaoka18:overcomplete}, is that, to generalize a compact basis from
the two- to the $n$-point GF, we must expand the semi-PGFs instead of the full GF \cite{Boehnke11}.

Indeed, replacing the kernels in Eq.~(\ref{eq:Giw-conv-prime}) 
with their truncated singular-value expansion (\ref{eq:sve}) yields:
\begin{equation}
\begin{split}
    &G^\prime_{\bar 1\ldots \bar n}(\ivv) \approx \beta\delta_{\vec\nu} 
    \sum_{\ell_1=0}^{L-1} \cdots \sum_{\ell_{n-1}=0}^{L-1}
     \\ &\qquad\times
     U_{\ell_1}\!(\iv_{\bar1})\cdots  U_{\ell_{n-1}}\!(\iv_{\bar1}+\ldots+\iv_{\overline{n-1}})\ g_{\bar 1\ldots\bar n,\vec\ell},
\end{split}
\label{eq:Giw-overcomplete}
\end{equation}
where $g$ are again a set of basis coefficients, given by:

\begin{align}
g_{\bar 1\ldots\bar n,\vec\ell}
    &= S_{\ell_1} \cdots S_{\ell_{n-1}} \rho^\prime_{\bar 1\ldots\bar n,\vec\ell},
    \label{eq:gIR} \\
\rho^\prime_{\bar 1\ldots\bar n,\vec\ell}
    &= \int d^{n-1}\vec{\epsilon}\, V_{\ell_1}(\epsilon_1)\cdots V_{\ell_{n-1}}(\epsilon_{n-1}) \rho'_{\bar1\ldots\bar n}(\vec\epsilon).
    \label{eq:rhoIR}
\end{align}

This is the IR of an arbitrary $n$-point GF. Since the basis coefficients (\ref{eq:gIR}) are multiplied by the quickly decaying singular values $S_\ell$, we need to store only $\bigO(nL^{n-1})$ coefficients
\footnote{Naively, one would expect $\bigO(n!L^{n-1})$ coefficients, however, the $n!$ is cancelled:
due to the exponential decay of the singular values, only a fraction $1/(n-1)!$ of the hypercube satisfies $S_{\ell_1}\cdots S_{\ell_{n-1}}>\varepsilon$~\cite{Wallerberger21:BSE}.}%
, where $L\sim \log(\beta\emax)\log(\varepsilon^{-1})$.
Equations~(\ref{eq:Giw-overcomplete}) to (\ref{eq:rhoIR}) were previously derived for the three- and four-point case~\cite{Shinaoka18:overcomplete}. Table~\ref{table:IRtoPerm_Group} relates the pairs of permutations indexing of the representations to the representation index used in Ref.~\cite{Wallerberger21:BSE}. Importantly, we now have a formula for arbitrary $n$.

A brief comment about bosonic arguments in Eq.~(\ref{eq:Giw-overcomplete}) is in order:
as alluded to in Sec.~\ref{sec:imagn}, whenever a sum of fermionic frequencies is exactly zero,
additional terms appear. One can show that these terms can be 
formally absorbed by augmenting the one-particle basis $U_\ell$ in Eq.~(\ref{eq:Giw-overcomplete})
\cite{Shinaoka18:overcomplete}. Numerically, this augmentation is usually not necessary
as the additional basis functions are almost linearly dependent on the other
basis functions \cite{Wallerberger21:BSE}.

We note that the construction (\ref{eq:Giw-overcomplete}) is in principle independent of the actual form of
the basis function used. In particular, one can replace the underlying IR basis by a finite sum over  
real \cite{Kaye22} or complex \cite{Ying22,zhang23arxiv} poles:
\begin{equation}
\begin{split}
    & G_{\bar 1\ldots \bar n}(\ivv) \approx \beta\delta_{\vec\nu} \\
    &\quad \times
    \sum_{p_1,\ldots,p_{n-1}} \frac{\rho_{\bar 1\ldots \bar n,p_1\ldots p_{n-1}}}%
    {(\iv_{\bar1} - \epsilon_{p_1})\cdots(\iv_{\bar1} + \ldots + \iv_{\overline{n-1}} - \epsilon_{p_{n-1}})},
\end{split}
\label{eq:Giw-conv-poles}
\end{equation}
where $\rho_{\bar 1\ldots \bar n,\vec p}$ and $\epsilon_p$ for $p = 1,\ldots,L$ are now parameters to be fitted, either directly for each semi-PGF or in the overcomplete sense (see below).

\begin{figure}
    \includegraphics[trim=1em 2em 4.5em 3.5em,clip,width=.8\columnwidth]{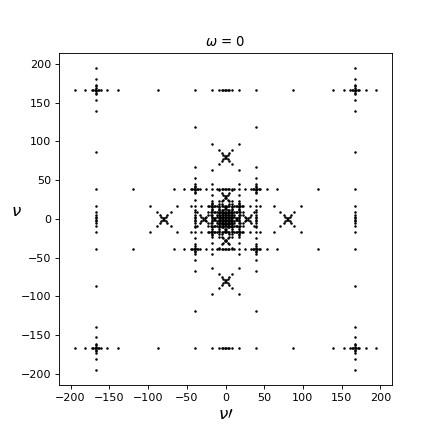}
    \caption{Sparse sampling points for the OIR basis, with $\beta = 100/D$, $\emax = D$, $\varepsilon = 10^{-3}$, plotted in $\omega = 0$ plane.}
    \label{fig:sampling_points}
\end{figure}

As in the two-point case, sparse sampling can be used to infer the basis coefficients (an example can be seen in Fig.~\ref{fig:sampling_points}). To this end, one
constructs a sampling frequency set $\mathcal V_n$ (e.g., by taking the direct product of $\mathcal V_2$ for
all the possible permutations) and turns Eqs.~(\ref{eq:Giw-conv}) and (\ref{eq:Giw-overcomplete}) into a least-squares problem:
\begin{equation}
\begin{split}
    \min_g &\sum_{\ivv\in\mathcal V_n} \bigg| G(\ivv) - \beta \primedsum_{\bar 1\ldots \bar n} 
    \sum_{\ell_1=0}^{L-1} \cdots \sum_{\ell_{n-1}=0}^{L-1}
     \\ &\times
     U_{\ell_1}\!(\iv_{\bar1})\cdots  U_{\ell_{n-1}}\!(\iv_{\bar1}+\ldots+\iv_{\overline{n-1}})\ g_{\bar 1\ldots\bar n,\vec\ell}\bigg|^2,
\end{split}
\label{eq:Giw-fit}
\end{equation}
which can be solved in $\bigO(n!L^{\lceil 2.5(n-1)\rceil})$ time~\cite{Wallerberger21:BSE}.

Given the basis coefficients $g$ in Eqs.~(\ref{eq:gIR}) and (\ref{eq:rhoIR}), we can invert these two equations to perform analytic continuation. However, we emphasize again that, since pairs of PSFs are combined in the imaginary-frequency GF [Eq.~(\ref{eq:rhoprime})], only semi-PSFs can be inferred.

Another, more practical problem is the following: since $U_\ell$ form a basis for the two-point GF, the basis expansion (\ref{eq:Giw-overcomplete}) is {\em overcomplete} for $L\to\infty$, since any one permutation already spans the full space, yet we sum over $n!/2$ permutations. For finite $L$, the basis functions formally do not have this issue, but they are still almost linearly dependent, which implies that the least-squares problem (\ref{eq:Giw-fit}) is ill-conditioned.
This in turn means that the basis coefficients $g$ strongly depend on the regularization scheme for the least-squares problem (\ref{eq:Giw-overcomplete}). This does not hinder the ability of the basis to compress, inter- and extrapolate imaginary-frequency data~\cite{Shinaoka18:overcomplete,Shinaoka20:tensornw,Wallerberger21:BSE}. However, it is a problem for analytic continuation, since it is unclear if the fitted coefficients have any connection
to the semi-PSFs. Exploring this is the subject of Sec.~\ref{sec:results}.

\section{Matching the overcomplete intermediate representation and partial Green's functions from data}
\label{sec:results}
In this section, we first compress imaginary-frequency data computed with NRG, and then compare the fitted functions (\ref{eq:g4iw-split}) with the exact expressions (\ref{eq:g4iw-exact}).

We use the particle-hole symmetric single-impurity Anderson model (SIAM) with a flat hybridization. Its Hamiltonian is:
\begin{equation}
\begin{split}
    H &= U \ddg_\uparrow \ddg_\downarrow \dee_\downarrow \dee_\uparrow
      - \tfrac12 U (\ddg_\uparrow \dee_\uparrow + \ddg_\downarrow \dee_\downarrow) \\
      & + V \sum_{p\sigma} (\cdag_{p\sigma} \dee_\sigma + \ddg_\sigma \cee_{p\sigma})
      + \sum_{p\sigma} \epsilon_p \cdag_{p\sigma} \cee_{p\sigma},
\end{split}
\label{eq:siam}
\end{equation}
where $\dee_\sigma$ and $\cee_{p\sigma}$ annihilate a spin-$\sigma$ electron on the impurity and in the bath with momentum $p$, respectively. Further, $U$ is the interaction strength, $V$ is the hopping amplitude between impurity and bath, taken to be constant, and $\epsilon_p$ is the energy of the corresponding bath level, taken uniformly distributed in the interval $\epsilon_p \in [-D,D]$.
We use $U = 0.2D$, temperature $\beta^{-1} = 10^{-2}D$, and a hybridization strength $\Delta=0.04D$,  defined as
\begin{equation}
    \sum_\epsilon \pi |V_\epsilon|^2 \delta(\omega -\epsilon) = \Delta\Theta(D-|\omega|).
\end{equation}

\subsection{Compression}
\label{sec:compression}

In the following, we illustrate the efficient compression of the Matsubara impurity two-particle GF. Its definition in imaginary time follows from Eq.~\eqref{eq:Gtau} with $(A_1, A_2, A_3, A_4) = (d_\uparrow, d_\uparrow{}^\dag, d_\uparrow, d_\uparrow{}^\dag)$.
In NRG, we compute this object as a sum of $4!$ PGFs, each of which is obtained by convolving a kernel and a PSF.

To specify on which Matsubara frequencies we store $G(\ivv)$, let us first define the particle--hole convention:
\begin{equation}
\begin{split}
    \nu_1, \nu_2, \nu_3, \nu_4  
    &\mapsto \nu=-\nu_2,  \nu'=\nu_3,  \omega=\nu_1+\nu_2,
       \label{eq:phconv}
\end{split}
\end{equation}
where $\nu$ and $\nu'$ are fermionic Matsubara frequencies and $\omega$ bosonic Matsubara frequencies.
We can now create a three dimensional box with axes $\nu, \nu'$, and $\omega$. We fill this frequency box with equidistant points in all dimensions. Each point is defined by a specific set of frequencies. The fermionic frequencies range from $-199\frac{\pi}{\beta}$ to $199\frac{\pi}{\beta}$ and the bosonic frequencies from $-200\frac{\pi}{\beta}$ to $200\frac{\pi}{\beta}$.

\begin{figure}
    \includegraphics[width=\columnwidth,trim=12 12 10 12]{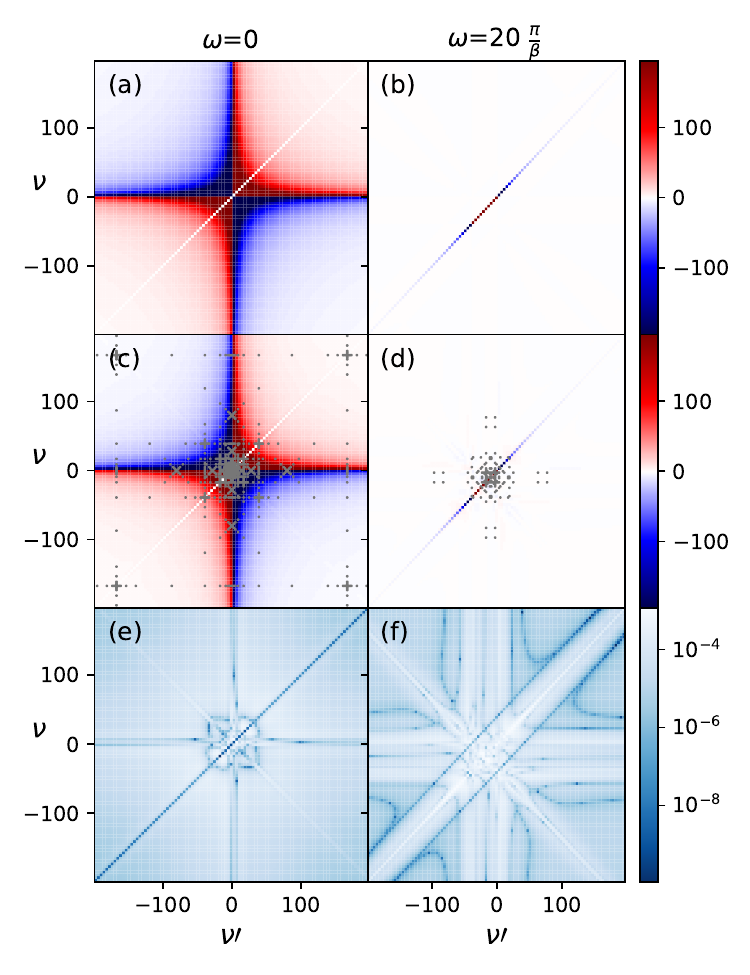}
    \caption{Four-point GF for the SIAM (\ref{eq:siam}) on Matsubara frequencies in the particle-hole convention for (a) $\omega=0$ and (b) $\omega=20\frac\pi\beta$. (c,d) The reconstructed data from the OIR fit for the same bosonic frequencies as in (a,b), with sampling frequencies indicated as gray dots.
    (e,f) Corresponding relative error.}
    \label{fig:error}
\end{figure}

Figure~\ref{fig:error}(a,b) shows $G(\ivv)$ from NRG in the $\nu,\nu'$ plane for (a) $\iw=0$ and (b) $\iw=20\frac\pi\beta$. There are non-trivial structures along the horizontal, vertical, and diagonal directions, which arise by summing the different PGFs. The diagonal of zero elements exemplifies why it is difficult to compress these objects. In fact, this figure was created with 8.04 million data points, occupying more than 6GB.

Using the OIR, we were able to compress these 6GB to 565 kB or about 0.2$\%$ of its original size.
For this, we estimate $\emax = D$ and choose $\varepsilon=10^{-3}$, which gives a linear basis size of $L=15$. The number of basis coefficients of the OIR is $11,952$.
To fit the data, we use Eq.~(\ref{eq:Giw-fit}), which for the four-point GF reads:
\begin{equation}
\begin{split}
    \min_g &\sum_{\ivv\in\mathcal V_4} \bigg| G(\ivv) - \primedsum_{\bar 1\bar 2\bar3\bar 4} 
    \sum_{\ell=0}^{L-1}  \sum_{m=0}^{L-1}
    \sum_{\ell'=0}^{L-1} 
     \\ &\times
     U_{\ell}(\iv_{\bar1}) U_{m}(\iv_{\bar1} + \iv_{\bar2}) U_{\ell'}(-\iv_{\bar4})\ g_{\bar 1\bar 2\bar3\bar 4,\ell m\ell'}\bigg|^2,
\end{split}
\label{eq:Giw-fit-4}
\end{equation}
where $\mathcal V_4$ are the sparse sampling points, indicated as gray dots in Fig.~\ref{fig:error}(c,d).
We write Eq.~(\ref{eq:Giw-fit-4}) as an ordinary least-squares problem with the loss function~\cite{Shinaoka18:overcomplete,Wallerberger21:BSE}:
\begin{equation}
    L = \| G-E \space g \|_2,
    \label{eq:loss}
\end{equation}
where $G$ is the target data, in our case obtained through NRG, and $E$ is the design matrix. We use an LSMR solver with loss function (\ref{eq:loss}) to obtain the coefficients $g$. The fitting procedures takes less than one minute on a six-core Ryzen 3600 CPU.

\begin{figure*}
\includegraphics[width=\textwidth]{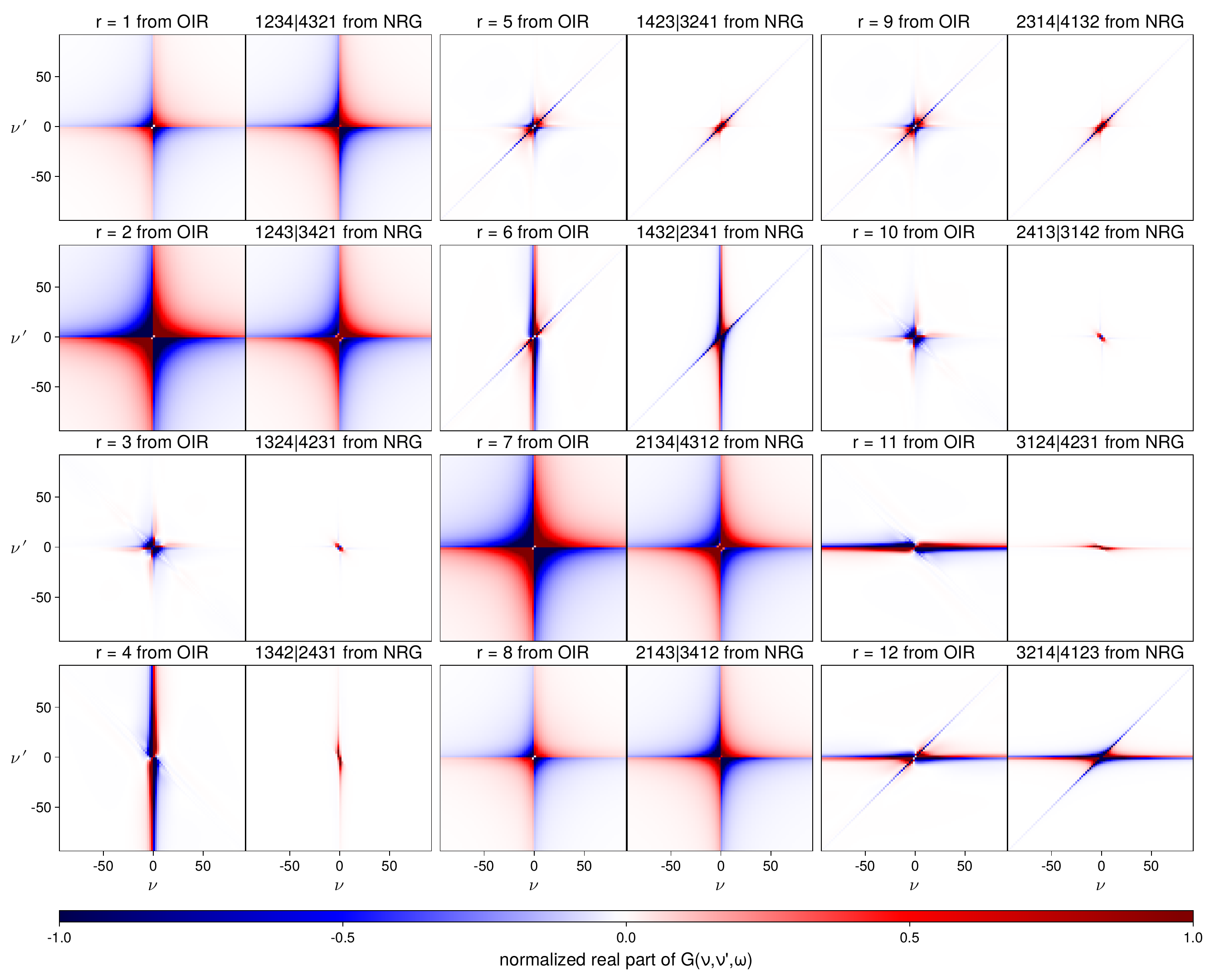}
\caption{
Comparison of semi-PGFs for the four-point GF of the SIAM. Each of the twelve pairs of panels depicts a semi-PGF when fitted from the OIR through Eq.~(\ref{eq:Giw-fit-4}) and then expanded using Eq.~(\ref{eq:g4iw-expand}) on the left side and compares it to the exact result from NRG (\ref{eq:g4iw-exact}) on the right side, cf. Table~\ref{table:IRtoPerm_Group}. The values are normalized by the maximum value per panel and plotted for $\omega = 0$ 
and $\nu $ and $\nu '$ ranging from -100 to 100 in the particle-hole convention (\ref{eq:phconv}).
}
\label{fig:channel}
\end{figure*}

To assess the error of this process, we first evaluate on the sparse sampling points by multiplying them with the corresponding design matrix $E$. This yields the predicted values of the GF on the sampling points, the so-called in-sample relative error, which was $9.95\times10^{-4}$, consistent with the accuracy goal of $\varepsilon=10^{-3}$. We also construct $E$ for the full frequency box of the NRG data with the sampling frequencies removed, yielding an out-of-sample relative error of $4.86\times10^{-3}$. The relative error
\begin{equation}
    |G(\ivv) - E(\ivv) g| / ||g||_\infty
    \label{eq:relerror}
\end{equation}
is plotted in Fig.~\ref{fig:error}(e,f), where the in-sample errors (locations of the dots) are below the accuracy goal, as expected, and the out-of-sample errors, away from the dots, are slightly higher but still comparable to the accuracy goal, indicating an absence of overfitting.
This means that we reduce the necessary frequency points from 8,040,000 points (which span the Matsubara box) to 19,282 sparse sampling points, while maintaining the same information up to the desired accuracy.

\subsection{Comparing partial Green's functions}

As outlined above, there is a two-to-one correspondence between the summands in the OIR and PGFs from NRG. Let us write this explicitly for the case of the two-particle GF, where the OIR (\ref{eq:Giw-overcomplete}) involves twelve semi-PGFs:
\begin{equation}
    G(\ivv) \approx \primedsum_{\bar 1\ldots \bar 4} G^\prime_{\bar 1\ldots \bar 4}(\ivv),
    \label{eq:g4iw-split}
\end{equation}
each of which is given by a basis expansion from a set of coefficients:
\begin{equation}
\begin{split}
    &G_{\bar 1\ldots \bar 4}(\ivv) = \beta\delta_{\vec\nu} \\
    &\quad\times \sum_{\ell\ell'm}
     U_{\ell}(\iv_{\bar1}) U_{m}(\iv_{\bar1}+\iv_{\bar2}) U_{\ell'}(-\iv_{\bar4}) g_{\bar 1\bar 2\bar 3\bar 4,\ell m\ell'}.
\end{split}
    \label{eq:g4iw-expand}
\end{equation}
In the OIR, the coefficients $g_{\bar 1\bar 2\bar 3\bar 4, \vec\ell}$ and hence the decomposition (\ref{eq:g4iw-split}) is fitted from imaginary-frequency data via a fitting problem similar to Eq.~(\ref{eq:Giw-fit}). This is enough to allow for a compressed representation and interpolation.

If the Hamiltonian is solved with, e.g., exact diagonalization \cite{Tanaka19} or NRG \cite{Lee21PRX}, then the ``true'' decomposition into semi-PGFs and, if desired, the ``true'' fitting coefficients obtained from the semi-PSFs (cf.\ Eqs.~\eqref{eq:gIR}, \eqref{eq:rhoIR}) can be computed:
\begin{equation}
    G'_{\bar 1\ldots \bar 4}(\ivv) = \beta\delta_{\vec\nu}
    \!\int\frac{d^3\vec{\epsilon}\ \rho'_{\bar1\bar2\bar3\bar4}(\epsilon_1,\epsilon_2,\epsilon_3)}%
    {(\iv_{\bar1} - \epsilon_1)(\iv_{\bar1} + \iv_{\bar2} -\epsilon_2)(-\iv_{\bar4} - \epsilon_3)}.
    \label{eq:g4iw-exact}
\end{equation}

Figure~\ref{fig:channel} compares the fitted and ``true'' semi-PGFs  for the SIAM (\ref{eq:siam}). For the OIR, the same fitting parameters were used as in Sec.~\ref{sec:compression}, with the exception of $\emax=2D+U=2.2$. Each of the twelve pairs of panels depicts a single summand of the OIR (\ref{eq:g4iw-expand}) on the left and the corresponding NRG semi-PGF on the right, cf.~Table~\ref{table:IRtoPerm_Group}. The comparison is made for bosonic frequency $\omega=0$, which exhibits the largest deviations, and plotted in the $\nu,\nu'$ plane in the particle-hole convention (\ref{eq:phconv}).
There is qualitative agreement for $r=1,2,6,7,8,12$  and discrepancy for $r=3,4,5,9,10, 11$.
This suggests that the loss function (\ref{eq:loss}), together with early stopping regularization performed by the LSMR, adversely affects the decomposition.

\subsection{Cosets}
\label{ssec:cosets}

\begin{figure}
\includegraphics[width=\columnwidth,trim=10 10 10 10]{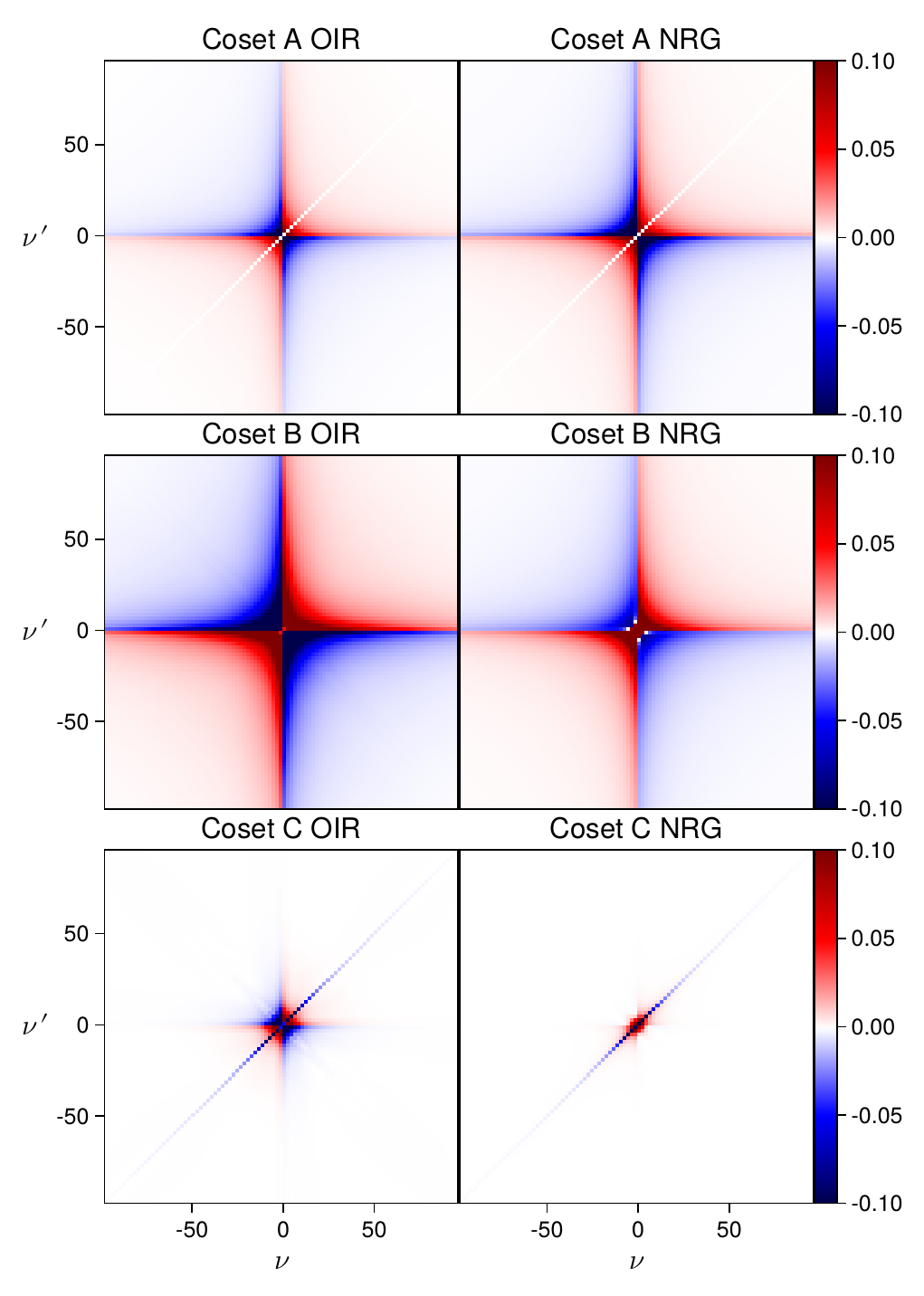}
\caption{
Comparison of 
PGFs between OIR and NRG, as in Fig.~\ref{fig:channel}, where each row now depicts the sum over a coset (group of representations as defined in Table~\ref{table:IRtoPerm_Group}).}
\label{fig:cosets}
\end{figure}

The OIR is, by design, overcomplete, so that a naive fitting problem is poorly conditioned. In other words, the fitting parameters are partially ambiguous.
One source of ambiguity is the relation between PSFs whose operator arguments are cyclic permutations of one another, see Eq.~(25) in Ref.~\cite{Kugler21}. For instance, the PSFs (in the present notation) for the permutations
1234 and 2341 obey:
\begin{equation}
\begin{split}
\rho_{2341}(\epsilon_2-\epsilon_1,\epsilon_3-\epsilon_1,-\epsilon_1) =
-\ee^{-\beta\epsilon_1} \rho_{1234}(\epsilon_1,\epsilon_2,\epsilon_3) .
\end{split}
\label{eq:cyclic}
\end{equation}
This partitions the permutations (and the corresponding PSFs) into three cosets, which cannot be related through either cyclic permutation or reversal of the arguments. We call these cosets A, B, C, and enumerate their elements in Table~\ref{table:IRtoPerm_Group}.

Figure~\ref{fig:cosets} shows the comparison of OIR and PGFs, decomposed only on the level of the cosets. As in Fig.~\ref{fig:channel}, we find merely partial qualitative agreement, albeit a somewhat better match.
This suggests that the cyclic permutation ambiguity is not the main source for the ill-conditioning of the fitting problem. We conjecture that this is due to the fact that the exponential factor in Eq.~(\ref{eq:cyclic}) may be poorly represented by the truncated IR expansion, which, if true, would lift the ambiguity.

\section{Conclusions}
\label{sec:conclusions}

We showed that, using the OIR, it is possible to compress data for  {two-particle Green's functions} computed from NRG and reconstruct it to the desired accuracy. For example, with an accuracy goal of $\varepsilon = 10^{-3}$, we achieved a data compression by a factor of 400  and an out-of-sample relative error of $4.86 \times 10^{-3}$. 

Further, we derived a two-to-one correspondence between the $n!$ PGFs, which can be obtained from exact diagonalization or NRG, and the $n!/2$ semi-PGF of the OIR.
For a two-particle ($n=4$) GF of the single-impurity Anderson model, we compared the 12 semi-PGFs of the OIR to their corresponding pair of NRG PGFs, and find a qualitative match but quantitative differences.
A further restriction of the 12 semi-PGF to only three cosets is possible, by grouping together terms corresponding to cyclic permutations.
Again, we find only a qualitative match between the OIR  and the original NRG data. 
Either the regularization used in the fitting procedure needs to be improved to better reflect the nature of the semi-PGFs or the overcompleteness of the OIR has to be mitigated for obtaining a better match.
 
Since the OIR fitting process introduces qualitative differences in the PGFs and analytic continuation is an ill-conditioned problem to begin with, it seems unlikely that 
the present scheme allows for an analytic continuation of two-particle GFs to real frequencies within reasonable error margins. Whether recent progress in understanding the analytic continuation of higher-order correlators~\cite{ge2023analytic} can, notwithstanding, help with the analytic continuation of  the OIR  is an interesting topic for future studies. 

\begin{acknowledgments}
We would like to thank Friedrich Krien for fruitful discussions. Further we acknowledge funding through the Austrian Science Fund (FWF) projects P 36332, P 36213, SFB Q-M\&S (FWF project ID F86), and Research Unit QUAST by the Deutsche Foschungsgemeinschaft (DFG; project ID FOR5249) and FWF (project ID I 5868). SSBL is supported by the New Faculty Startup Fund from Seoul National University, and also by the National Research Foundation of Korea (NRF) grant funded by the Korean government (MSIT) (No.~RS-2023-00214464).

For NRG calculations, the authors gratefully acknowledge the \href{www.gauss-centre.eu}{Gauss Centre for Supercomputing e.V.} for funding this project by providing computing time on the GCS Supercomputer SuperMUC-NG at \href{www.lrz.de}{Leibniz Supercomputing Centre}.
The Flatiron Institute is a division of the Simons Foundation. 

IR calculations were performed using the sparse-ir library \cite{SparseIR}. Codes for computing
the OIR and NRG data are available from the authors upon request and are forthcoming as open-source packages.
The data of our calculations for OIR and NRG is available upon request.

\end{acknowledgments}

\appendix

\bibliography{main}

\begin{thebibliography}{30}%
\makeatletter
\providecommand \@ifxundefined [1]{%
 \@ifx{#1\undefined}
}%
\providecommand \@ifnum [1]{%
 \ifnum #1\expandafter \@firstoftwo
 \else \expandafter \@secondoftwo
 \fi
}%
\providecommand \@ifx [1]{%
 \ifx #1\expandafter \@firstoftwo
 \else \expandafter \@secondoftwo
 \fi
}%
\providecommand \natexlab [1]{#1}%
\providecommand \enquote  [1]{``#1''}%
\providecommand \bibnamefont  [1]{#1}%
\providecommand \bibfnamefont [1]{#1}%
\providecommand \citenamefont [1]{#1}%
\providecommand \href@noop [0]{\@secondoftwo}%
\providecommand \href [0]{\begingroup \@sanitize@url \@href}%
\providecommand \@href[1]{\@@startlink{#1}\@@href}%
\providecommand \@@href[1]{\endgroup#1\@@endlink}%
\providecommand \@sanitize@url [0]{\catcode `\\12\catcode `\$12\catcode
  `\&12\catcode `\#12\catcode `\^12\catcode `\_12\catcode `\%12\relax}%
\providecommand \@@startlink[1]{}%
\providecommand \@@endlink[0]{}%
\providecommand \url  [0]{\begingroup\@sanitize@url \@url }%
\providecommand \@url [1]{\endgroup\@href {#1}{\urlprefix }}%
\providecommand \urlprefix  [0]{URL }%
\providecommand \Eprint [0]{\href }%
\providecommand \doibase [0]{https://doi.org/}%
\providecommand \selectlanguage [0]{\@gobble}%
\providecommand \bibinfo  [0]{\@secondoftwo}%
\providecommand \bibfield  [0]{\@secondoftwo}%
\providecommand \translation [1]{[#1]}%
\providecommand \BibitemOpen [0]{}%
\providecommand \bibitemStop [0]{}%
\providecommand \bibitemNoStop [0]{.\EOS\space}%
\providecommand \EOS [0]{\spacefactor3000\relax}%
\providecommand \BibitemShut  [1]{\csname bibitem#1\endcsname}%
\let\auto@bib@innerbib\@empty
\bibitem [{\citenamefont {Economou}(2006)}]{Economou2006}%
  \BibitemOpen
  \bibfield  {author} {\bibinfo {author} {\bibfnamefont {E.~N.}\ \bibnamefont
  {Economou}},\ }\href {https://doi.org/10.1007/3-540-28841-4} {\emph {\bibinfo
  {title} {Green’s Functions in Quantum Physics}}}\ (\bibinfo  {publisher}
  {Springer Berlin Heidelberg},\ \bibinfo {year} {2006})\BibitemShut {NoStop}%
\bibitem [{\citenamefont {Shinaoka}\ \emph {et~al.}(2017)\citenamefont
  {Shinaoka}, \citenamefont {Otsuki}, \citenamefont {Ohzeki},\ and\
  \citenamefont {Yoshimi}}]{Shinaoka17:compressing}%
  \BibitemOpen
  \bibfield  {author} {\bibinfo {author} {\bibfnamefont {H.}~\bibnamefont
  {Shinaoka}}, \bibinfo {author} {\bibfnamefont {J.}~\bibnamefont {Otsuki}},
  \bibinfo {author} {\bibfnamefont {M.}~\bibnamefont {Ohzeki}},\ and\ \bibinfo
  {author} {\bibfnamefont {K.}~\bibnamefont {Yoshimi}},\ }\bibfield  {title}
  {\bibinfo {title} {Compressing {Green's} function using intermediate
  representation between imaginary-time and real-frequency domains},\ }\href
  {https://doi.org/10.1103/PhysRevB.96.035147} {\bibfield  {journal} {\bibinfo
  {journal} {Phys. Rev. B}\ }\textbf {\bibinfo {volume} {96}},\ \bibinfo
  {pages} {35147} (\bibinfo {year} {2017})}\BibitemShut {NoStop}%
\bibitem [{\citenamefont {Li}\ \emph {et~al.}(2020)\citenamefont {Li},
  \citenamefont {Wallerberger}, \citenamefont {Chikano}, \citenamefont {Yeh},
  \citenamefont {Gull},\ and\ \citenamefont {Shinaoka}}]{Li20}%
  \BibitemOpen
  \bibfield  {author} {\bibinfo {author} {\bibfnamefont {J.}~\bibnamefont
  {Li}}, \bibinfo {author} {\bibfnamefont {M.}~\bibnamefont {Wallerberger}},
  \bibinfo {author} {\bibfnamefont {N.}~\bibnamefont {Chikano}}, \bibinfo
  {author} {\bibfnamefont {C.-N.}\ \bibnamefont {Yeh}}, \bibinfo {author}
  {\bibfnamefont {E.}~\bibnamefont {Gull}},\ and\ \bibinfo {author}
  {\bibfnamefont {H.}~\bibnamefont {Shinaoka}},\ }\bibfield  {title} {\bibinfo
  {title} {Sparse sampling approach to efficient ab initio calculations at
  finite temperature},\ }\href {https://doi.org/10.1103/physrevb.101.035144}
  {\bibfield  {journal} {\bibinfo  {journal} {Phys. Rev. B}\ }\textbf {\bibinfo
  {volume} {101}},\ \bibinfo {pages} {035144} (\bibinfo {year}
  {2020})}\BibitemShut {NoStop}%
\bibitem [{\citenamefont {Shinaoka}\ \emph {et~al.}(2023)\citenamefont
  {Shinaoka}, \citenamefont {Wallerberger}, \citenamefont {Murakami},
  \citenamefont {Nogaki}, \citenamefont {Sakurai}, \citenamefont {Werner},\
  and\ \citenamefont {Kauch}}]{ShinaokaPRX23}%
  \BibitemOpen
  \bibfield  {author} {\bibinfo {author} {\bibfnamefont {H.}~\bibnamefont
  {Shinaoka}}, \bibinfo {author} {\bibfnamefont {M.}~\bibnamefont
  {Wallerberger}}, \bibinfo {author} {\bibfnamefont {Y.}~\bibnamefont
  {Murakami}}, \bibinfo {author} {\bibfnamefont {K.}~\bibnamefont {Nogaki}},
  \bibinfo {author} {\bibfnamefont {R.}~\bibnamefont {Sakurai}}, \bibinfo
  {author} {\bibfnamefont {P.}~\bibnamefont {Werner}},\ and\ \bibinfo {author}
  {\bibfnamefont {A.}~\bibnamefont {Kauch}},\ }\bibfield  {title} {\bibinfo
  {title} {Multiscale space-time ansatz for correlation functions of quantum
  systems based on quantics tensor trains},\ }\href
  {https://doi.org/10.1103/PhysRevX.13.021015} {\bibfield  {journal} {\bibinfo
  {journal} {Phys. Rev. X}\ }\textbf {\bibinfo {volume} {13}},\ \bibinfo
  {pages} {021015} (\bibinfo {year} {2023})}\BibitemShut {NoStop}%
\bibitem [{\citenamefont {Ritter}\ \emph {et~al.}(2024)\citenamefont {Ritter},
  \citenamefont {N\'u\~nez Fern\'andez}, \citenamefont {Wallerberger},
  \citenamefont {von Delft}, \citenamefont {Shinaoka},\ and\ \citenamefont
  {Waintal}}]{Ritter23}%
  \BibitemOpen
  \bibfield  {author} {\bibinfo {author} {\bibfnamefont {M.~K.}\ \bibnamefont
  {Ritter}}, \bibinfo {author} {\bibfnamefont {Y.}~\bibnamefont {N\'u\~nez
  Fern\'andez}}, \bibinfo {author} {\bibfnamefont {M.}~\bibnamefont
  {Wallerberger}}, \bibinfo {author} {\bibfnamefont {J.}~\bibnamefont {von
  Delft}}, \bibinfo {author} {\bibfnamefont {H.}~\bibnamefont {Shinaoka}},\
  and\ \bibinfo {author} {\bibfnamefont {X.}~\bibnamefont {Waintal}},\
  }\bibfield  {title} {\bibinfo {title} {Quantics tensor cross interpolation
  for high-resolution parsimonious representations of multivariate functions},\
  }\href {https://doi.org/10.1103/PhysRevLett.132.056501} {\bibfield  {journal}
  {\bibinfo  {journal} {Phys. Rev. Lett.}\ }\textbf {\bibinfo {volume} {132}},\
  \bibinfo {pages} {056501} (\bibinfo {year} {2024})}\BibitemShut {NoStop}%
\bibitem [{\citenamefont {N\'u\~nez Fern\'andez}\ \emph
  {et~al.}(2022)\citenamefont {N\'u\~nez Fern\'andez}, \citenamefont {Jeannin},
  \citenamefont {Dumitrescu}, \citenamefont {Kloss}, \citenamefont {Kaye},
  \citenamefont {Parcollet},\ and\ \citenamefont {Waintal}}]{FernandezPRX22}%
  \BibitemOpen
  \bibfield  {author} {\bibinfo {author} {\bibfnamefont {Y.}~\bibnamefont
  {N\'u\~nez Fern\'andez}}, \bibinfo {author} {\bibfnamefont {M.}~\bibnamefont
  {Jeannin}}, \bibinfo {author} {\bibfnamefont {P.~T.}\ \bibnamefont
  {Dumitrescu}}, \bibinfo {author} {\bibfnamefont {T.}~\bibnamefont {Kloss}},
  \bibinfo {author} {\bibfnamefont {J.}~\bibnamefont {Kaye}}, \bibinfo {author}
  {\bibfnamefont {O.}~\bibnamefont {Parcollet}},\ and\ \bibinfo {author}
  {\bibfnamefont {X.}~\bibnamefont {Waintal}},\ }\bibfield  {title} {\bibinfo
  {title} {Learning {F}eynman diagrams with tensor trains},\ }\href
  {https://doi.org/10.1103/PhysRevX.12.041018} {\bibfield  {journal} {\bibinfo
  {journal} {Phys. Rev. X}\ }\textbf {\bibinfo {volume} {12}},\ \bibinfo
  {pages} {041018} (\bibinfo {year} {2022})}\BibitemShut {NoStop}%
\bibitem [{\citenamefont {Wallerberger}(2016)}]{Wallerberger2016}%
  \BibitemOpen
  \bibfield  {author} {\bibinfo {author} {\bibfnamefont {M.}~\bibnamefont
  {Wallerberger}},\ }\emph {\bibinfo {title} {{w2dynamics}: {Continuous} time
  quantum {Monte Carlo} calculations of one-and two-particle propagators}},\
  \href {http://repositum.tuwien.ac.at/urn:nbn:at:at-ubtuw:1-3537} {Ph.D.
  thesis},\ \bibinfo  {school} {Technische Universit\"at Wien} (\bibinfo {year}
  {2016})\BibitemShut {NoStop}%
\bibitem [{\citenamefont {Li}\ \emph {et~al.}(2016)\citenamefont {Li},
  \citenamefont {Wentzell}, \citenamefont {Pudleiner}, \citenamefont
  {Thunstr\"om},\ and\ \citenamefont {Held}}]{GangLi16}%
  \BibitemOpen
  \bibfield  {author} {\bibinfo {author} {\bibfnamefont {G.}~\bibnamefont
  {Li}}, \bibinfo {author} {\bibfnamefont {N.}~\bibnamefont {Wentzell}},
  \bibinfo {author} {\bibfnamefont {P.}~\bibnamefont {Pudleiner}}, \bibinfo
  {author} {\bibfnamefont {P.}~\bibnamefont {Thunstr\"om}},\ and\ \bibinfo
  {author} {\bibfnamefont {K.}~\bibnamefont {Held}},\ }\bibfield  {title}
  {\bibinfo {title} {Efficient implementation of the parquet equations: Role of
  the reducible vertex function and its kernel approximation},\ }\href
  {https://doi.org/10.1103/PhysRevB.93.165103} {\bibfield  {journal} {\bibinfo
  {journal} {Phys. Rev. B}\ }\textbf {\bibinfo {volume} {93}},\ \bibinfo
  {pages} {165103} (\bibinfo {year} {2016})}\BibitemShut {NoStop}%
\bibitem [{\citenamefont {Wentzell}\ \emph {et~al.}(2020)\citenamefont
  {Wentzell}, \citenamefont {Li}, \citenamefont {Tagliavini}, \citenamefont
  {Taranto}, \citenamefont {Rohringer}, \citenamefont {Held}, \citenamefont
  {Toschi},\ and\ \citenamefont {Andergassen}}]{Wentzell20}%
  \BibitemOpen
  \bibfield  {author} {\bibinfo {author} {\bibfnamefont {N.}~\bibnamefont
  {Wentzell}}, \bibinfo {author} {\bibfnamefont {G.}~\bibnamefont {Li}},
  \bibinfo {author} {\bibfnamefont {A.}~\bibnamefont {Tagliavini}}, \bibinfo
  {author} {\bibfnamefont {C.}~\bibnamefont {Taranto}}, \bibinfo {author}
  {\bibfnamefont {G.}~\bibnamefont {Rohringer}}, \bibinfo {author}
  {\bibfnamefont {K.}~\bibnamefont {Held}}, \bibinfo {author} {\bibfnamefont
  {A.}~\bibnamefont {Toschi}},\ and\ \bibinfo {author} {\bibfnamefont
  {S.}~\bibnamefont {Andergassen}},\ }\bibfield  {title} {\bibinfo {title}
  {High-frequency asymptotics of the vertex function: Diagrammatic
  parametrization and algorithmic implementation},\ }\href
  {https://doi.org/10.1103/PhysRevB.102.085106} {\bibfield  {journal} {\bibinfo
   {journal} {Phys. Rev. B}\ }\textbf {\bibinfo {volume} {102}},\ \bibinfo
  {pages} {085106} (\bibinfo {year} {2020})}\BibitemShut {NoStop}%
\bibitem [{\citenamefont {Krien}\ \emph {et~al.}(2021)\citenamefont {Krien},
  \citenamefont {Kauch},\ and\ \citenamefont {Held}}]{Krien2021}%
  \BibitemOpen
  \bibfield  {author} {\bibinfo {author} {\bibfnamefont {F.}~\bibnamefont
  {Krien}}, \bibinfo {author} {\bibfnamefont {A.}~\bibnamefont {Kauch}},\ and\
  \bibinfo {author} {\bibfnamefont {K.}~\bibnamefont {Held}},\ }\bibfield
  {title} {\bibinfo {title} {Tiling with triangles: parquet and {$GW\gamma$}
  methods unified},\ }\href {https://doi.org/10.1103/PhysRevResearch.3.013149}
  {\bibfield  {journal} {\bibinfo  {journal} {Phys. Rev. Res.}\ }\textbf
  {\bibinfo {volume} {3}},\ \bibinfo {pages} {013149} (\bibinfo {year}
  {2021})}\BibitemShut {NoStop}%
\bibitem [{\citenamefont {Prokof'ev}\ and\ \citenamefont
  {Svistunov}(2008)}]{Prokofev08}%
  \BibitemOpen
  \bibfield  {author} {\bibinfo {author} {\bibfnamefont {N.}~\bibnamefont
  {Prokof'ev}}\ and\ \bibinfo {author} {\bibfnamefont {B.}~\bibnamefont
  {Svistunov}},\ }\bibfield  {title} {\bibinfo {title} {{Fermi}-polaron
  problem: Diagrammatic {Monte Carlo} method for divergent sign-alternating
  series},\ }\href {https://doi.org/10.1103/PhysRevB.77.020408} {\bibfield
  {journal} {\bibinfo  {journal} {Phys. Rev. B}\ }\textbf {\bibinfo {volume}
  {77}},\ \bibinfo {pages} {020408} (\bibinfo {year} {2008})}\BibitemShut
  {NoStop}%
\bibitem [{\citenamefont {Taheridehkordi}\ \emph {et~al.}(2019)\citenamefont
  {Taheridehkordi}, \citenamefont {Curnoe},\ and\ \citenamefont
  {LeBlanc}}]{Taheridehkordi19}%
  \BibitemOpen
  \bibfield  {author} {\bibinfo {author} {\bibfnamefont {A.}~\bibnamefont
  {Taheridehkordi}}, \bibinfo {author} {\bibfnamefont {S.~H.}\ \bibnamefont
  {Curnoe}},\ and\ \bibinfo {author} {\bibfnamefont {J.~P.~F.}\ \bibnamefont
  {LeBlanc}},\ }\bibfield  {title} {\bibinfo {title} {Algorithmic {Matsubara}
  integration for {Hubbard}-like models},\ }\href
  {https://doi.org/10.1103/PhysRevB.99.035120} {\bibfield  {journal} {\bibinfo
  {journal} {Phys. Rev. B}\ }\textbf {\bibinfo {volume} {99}},\ \bibinfo
  {pages} {035120} (\bibinfo {year} {2019})}\BibitemShut {NoStop}%
\bibitem [{\citenamefont {Shinaoka}\ \emph {et~al.}(2018)\citenamefont
  {Shinaoka}, \citenamefont {Otsuki}, \citenamefont {Haule}, \citenamefont
  {Wallerberger}, \citenamefont {Gull}, \citenamefont {Yoshimi},\ and\
  \citenamefont {Ohzeki}}]{Shinaoka18:overcomplete}%
  \BibitemOpen
  \bibfield  {author} {\bibinfo {author} {\bibfnamefont {H.}~\bibnamefont
  {Shinaoka}}, \bibinfo {author} {\bibfnamefont {J.}~\bibnamefont {Otsuki}},
  \bibinfo {author} {\bibfnamefont {K.}~\bibnamefont {Haule}}, \bibinfo
  {author} {\bibfnamefont {M.}~\bibnamefont {Wallerberger}}, \bibinfo {author}
  {\bibfnamefont {E.}~\bibnamefont {Gull}}, \bibinfo {author} {\bibfnamefont
  {K.}~\bibnamefont {Yoshimi}},\ and\ \bibinfo {author} {\bibfnamefont
  {M.}~\bibnamefont {Ohzeki}},\ }\bibfield  {title} {\bibinfo {title}
  {Overcomplete compact representation of two-particle {Green's} functions},\
  }\href {https://doi.org/10.1103/PhysRevB.97.205111} {\bibfield  {journal}
  {\bibinfo  {journal} {Phys. Rev. B}\ }\textbf {\bibinfo {volume} {97}},\
  \bibinfo {pages} {205111} (\bibinfo {year} {2018})}\BibitemShut {NoStop}%
\bibitem [{\citenamefont {Wallerberger}\ \emph {et~al.}(2021)\citenamefont
  {Wallerberger}, \citenamefont {Shinaoka},\ and\ \citenamefont
  {Kauch}}]{Wallerberger21:BSE}%
  \BibitemOpen
  \bibfield  {author} {\bibinfo {author} {\bibfnamefont {M.}~\bibnamefont
  {Wallerberger}}, \bibinfo {author} {\bibfnamefont {H.}~\bibnamefont
  {Shinaoka}},\ and\ \bibinfo {author} {\bibfnamefont {A.}~\bibnamefont
  {Kauch}},\ }\bibfield  {title} {\bibinfo {title} {Solving the
  {Bethe--Salpeter} equation with exponential convergence},\ }\href
  {https://doi.org/10.1103/PhysRevResearch.3.033168} {\bibfield  {journal}
  {\bibinfo  {journal} {Phys. Rev. Research}\ }\textbf {\bibinfo {volume}
  {3}},\ \bibinfo {pages} {033168} (\bibinfo {year} {2021})}\BibitemShut
  {NoStop}%
\bibitem [{\citenamefont {Kugler}\ \emph {et~al.}(2021)\citenamefont {Kugler},
  \citenamefont {Lee},\ and\ \citenamefont {von Delft}}]{Kugler21}%
  \BibitemOpen
  \bibfield  {author} {\bibinfo {author} {\bibfnamefont {F.~B.}\ \bibnamefont
  {Kugler}}, \bibinfo {author} {\bibfnamefont {S.-S.~B.}\ \bibnamefont {Lee}},\
  and\ \bibinfo {author} {\bibfnamefont {J.}~\bibnamefont {von Delft}},\
  }\bibfield  {title} {\bibinfo {title} {Multipoint correlation functions:
  Spectral representation and numerical evaluation},\ }\href
  {https://doi.org/10.1103/PhysRevX.11.041006} {\bibfield  {journal} {\bibinfo
  {journal} {Phys. Rev. X}\ }\textbf {\bibinfo {volume} {11}},\ \bibinfo
  {pages} {041006} (\bibinfo {year} {2021})}\BibitemShut {NoStop}%
\bibitem [{\citenamefont {Tanaka}(2019)}]{Tanaka19}%
  \BibitemOpen
  \bibfield  {author} {\bibinfo {author} {\bibfnamefont {A.}~\bibnamefont
  {Tanaka}},\ }\bibfield  {title} {\bibinfo {title} {Metal-insulator transition
  in the two-di\-men\-sion\-al {Hubbard} model: Dual fermion approach with
  {Lanczos} exact diagonalization},\ }\href
  {https://doi.org/10.1103/PhysRevB.99.205133} {\bibfield  {journal} {\bibinfo
  {journal} {Phys. Rev. B}\ }\textbf {\bibinfo {volume} {99}},\ \bibinfo
  {pages} {205133} (\bibinfo {year} {2019})}\BibitemShut {NoStop}%
\bibitem [{\citenamefont {Lee}\ \emph {et~al.}(2021)\citenamefont {Lee},
  \citenamefont {Kugler},\ and\ \citenamefont {von Delft}}]{Lee21PRX}%
  \BibitemOpen
  \bibfield  {author} {\bibinfo {author} {\bibfnamefont {S.-S.~B.}\
  \bibnamefont {Lee}}, \bibinfo {author} {\bibfnamefont {F.~B.}\ \bibnamefont
  {Kugler}},\ and\ \bibinfo {author} {\bibfnamefont {J.}~\bibnamefont {von
  Delft}},\ }\bibfield  {title} {\bibinfo {title} {Computing local multipoint
  correlators using the numerical renormalization group},\ }\href
  {https://doi.org/10.1103/PhysRevX.11.041007} {\bibfield  {journal} {\bibinfo
  {journal} {Phys. Rev. X}\ }\textbf {\bibinfo {volume} {11}},\ \bibinfo
  {pages} {041007} (\bibinfo {year} {2021})}\BibitemShut {NoStop}%
\bibitem [{\citenamefont {Halbinger}\ \emph {et~al.}(2023)\citenamefont
  {Halbinger}, \citenamefont {Schneider},\ and\ \citenamefont
  {Sbierski}}]{Halbinger23}%
  \BibitemOpen
  \bibfield  {author} {\bibinfo {author} {\bibfnamefont {J.}~\bibnamefont
  {Halbinger}}, \bibinfo {author} {\bibfnamefont {B.}~\bibnamefont
  {Schneider}},\ and\ \bibinfo {author} {\bibfnamefont {B.}~\bibnamefont
  {Sbierski}},\ }\bibfield  {title} {\bibinfo {title} {{Spectral representation
  of Matsubara n-point functions: Exact kernel functions and applications}},\
  }\href {https://doi.org/10.21468/SciPostPhys.15.5.183} {\bibfield  {journal}
  {\bibinfo  {journal} {SciPost Phys.}\ }\textbf {\bibinfo {volume} {15}},\
  \bibinfo {pages} {183} (\bibinfo {year} {2023})}\BibitemShut {NoStop}%
\bibitem [{\citenamefont {Bryan}(1990)}]{Bryan90}%
  \BibitemOpen
  \bibfield  {author} {\bibinfo {author} {\bibfnamefont {R.~K.}\ \bibnamefont
  {Bryan}},\ }\bibfield  {title} {\bibinfo {title} {Maximum entropy analysis of
  oversampled data problems},\ }\href {https://doi.org/10.1007/BF02427376}
  {\bibfield  {journal} {\bibinfo  {journal} {Eur. Biophys. J.}\ }\textbf
  {\bibinfo {volume} {18}},\ \bibinfo {pages} {165} (\bibinfo {year}
  {1990})}\BibitemShut {NoStop}%
\bibitem [{\citenamefont {Otsuki}\ \emph {et~al.}(2017)\citenamefont {Otsuki},
  \citenamefont {Ohzeki}, \citenamefont {Shinaoka},\ and\ \citenamefont
  {Yoshimi}}]{Otsuki17}%
  \BibitemOpen
  \bibfield  {author} {\bibinfo {author} {\bibfnamefont {J.}~\bibnamefont
  {Otsuki}}, \bibinfo {author} {\bibfnamefont {M.}~\bibnamefont {Ohzeki}},
  \bibinfo {author} {\bibfnamefont {H.}~\bibnamefont {Shinaoka}},\ and\
  \bibinfo {author} {\bibfnamefont {K.}~\bibnamefont {Yoshimi}},\ }\bibfield
  {title} {\bibinfo {title} {Sparse modeling approach to analytical
  continuation of imaginary-time quantum {Monte Carlo} data},\ }\href
  {https://doi.org/10.1103/PhysRevE.95.061302} {\bibfield  {journal} {\bibinfo
  {journal} {Phys. Rev. E}\ }\textbf {\bibinfo {volume} {95}},\ \bibinfo
  {pages} {061302} (\bibinfo {year} {2017})}\BibitemShut {NoStop}%
\bibitem [{\citenamefont {Chikano}\ \emph {et~al.}(2018)\citenamefont
  {Chikano}, \citenamefont {Otsuki},\ and\ \citenamefont
  {Shinaoka}}]{Chikano18}%
  \BibitemOpen
  \bibfield  {author} {\bibinfo {author} {\bibfnamefont {N.}~\bibnamefont
  {Chikano}}, \bibinfo {author} {\bibfnamefont {J.}~\bibnamefont {Otsuki}},\
  and\ \bibinfo {author} {\bibfnamefont {H.}~\bibnamefont {Shinaoka}},\
  }\bibfield  {title} {\bibinfo {title} {Performance analysis of a physically
  constructed orthogonal representation of imaginary-time {Green's} function},\
  }\href {https://doi.org/10.1103/PhysRevB.98.035104} {\bibfield  {journal}
  {\bibinfo  {journal} {Phys. Rev. B}\ }\textbf {\bibinfo {volume} {98}},\
  \bibinfo {pages} {035104} (\bibinfo {year} {2018})}\BibitemShut {NoStop}%
\bibitem [{\citenamefont {Shinaoka}\ \emph {et~al.}(2022)\citenamefont
  {Shinaoka}, \citenamefont {Chikano}, \citenamefont {Gull}, \citenamefont
  {Li}, \citenamefont {Nomoto}, \citenamefont {Otsuki}, \citenamefont
  {Wallerberger}, \citenamefont {Wang},\ and\ \citenamefont
  {Yoshimi}}]{SciPost21}%
  \BibitemOpen
  \bibfield  {author} {\bibinfo {author} {\bibfnamefont {H.}~\bibnamefont
  {Shinaoka}}, \bibinfo {author} {\bibfnamefont {N.}~\bibnamefont {Chikano}},
  \bibinfo {author} {\bibfnamefont {E.}~\bibnamefont {Gull}}, \bibinfo {author}
  {\bibfnamefont {J.}~\bibnamefont {Li}}, \bibinfo {author} {\bibfnamefont
  {T.}~\bibnamefont {Nomoto}}, \bibinfo {author} {\bibfnamefont
  {J.}~\bibnamefont {Otsuki}}, \bibinfo {author} {\bibfnamefont
  {M.}~\bibnamefont {Wallerberger}}, \bibinfo {author} {\bibfnamefont
  {T.}~\bibnamefont {Wang}},\ and\ \bibinfo {author} {\bibfnamefont
  {K.}~\bibnamefont {Yoshimi}},\ }\bibfield  {title} {\bibinfo {title}
  {Efficient ab initio many-body calculations based on sparse modeling of
  {Matsubara Green's} function},\ }\href
  {https://doi.org/10.21468/SciPostPhysLectNotes.63} {\bibfield  {journal}
  {\bibinfo  {journal} {SciPost Phys. Lect. Notes}\ ,\ \bibinfo {pages} {63}}
  (\bibinfo {year} {2022})}\BibitemShut {NoStop}%
\bibitem [{\citenamefont {Boehnke}\ \emph {et~al.}(2011)\citenamefont
  {Boehnke}, \citenamefont {Hafermann}, \citenamefont {Ferrero}, \citenamefont
  {Lechermann},\ and\ \citenamefont {Parcollet}}]{Boehnke11}%
  \BibitemOpen
  \bibfield  {author} {\bibinfo {author} {\bibfnamefont {L.}~\bibnamefont
  {Boehnke}}, \bibinfo {author} {\bibfnamefont {H.}~\bibnamefont {Hafermann}},
  \bibinfo {author} {\bibfnamefont {M.}~\bibnamefont {Ferrero}}, \bibinfo
  {author} {\bibfnamefont {F.}~\bibnamefont {Lechermann}},\ and\ \bibinfo
  {author} {\bibfnamefont {O.}~\bibnamefont {Parcollet}},\ }\bibfield  {title}
  {\bibinfo {title} {Orthogonal polynomial representation of imaginary-time
  green's functions},\ }\href {https://doi.org/10.1103/PhysRevB.84.075145}
  {\bibfield  {journal} {\bibinfo  {journal} {Phys. Rev. B}\ }\textbf {\bibinfo
  {volume} {84}},\ \bibinfo {pages} {075145} (\bibinfo {year}
  {2011})}\BibitemShut {NoStop}%
\bibitem [{Note1()}]{Note1}%
  \BibitemOpen
  \bibinfo {note} {Naively, one would expect $\protect \mathcal {O}(n!L^{n-1})$
  coefficients, however, the $n!$ is cancelled: due to the exponential decay of
  the singular values, only a fraction $1/(n-1)!$ of the hypercube satisfies
  $S_{\ell _1}\protect \cdots S_{\ell _{n-1}}>\varepsilon $~\cite
  {Wallerberger21:BSE}.}\BibitemShut {Stop}%
\bibitem [{\citenamefont {Kaye}\ \emph {et~al.}(2022)\citenamefont {Kaye},
  \citenamefont {Chen},\ and\ \citenamefont {Parcollet}}]{Kaye22}%
  \BibitemOpen
  \bibfield  {author} {\bibinfo {author} {\bibfnamefont {J.}~\bibnamefont
  {Kaye}}, \bibinfo {author} {\bibfnamefont {K.}~\bibnamefont {Chen}},\ and\
  \bibinfo {author} {\bibfnamefont {O.}~\bibnamefont {Parcollet}},\ }\bibfield
  {title} {\bibinfo {title} {Discrete lehmann representation of imaginary time
  green's functions},\ }\href {https://doi.org/10.1103/PhysRevB.105.235115}
  {\bibfield  {journal} {\bibinfo  {journal} {Phys. Rev. B}\ }\textbf {\bibinfo
  {volume} {105}},\ \bibinfo {pages} {235115} (\bibinfo {year}
  {2022})}\BibitemShut {NoStop}%
\bibitem [{\citenamefont {Ying}(2022)}]{Ying22}%
  \BibitemOpen
  \bibfield  {author} {\bibinfo {author} {\bibfnamefont {L.}~\bibnamefont
  {Ying}},\ }\bibfield  {title} {\bibinfo {title} {Pole recovery from noisy
  data on imaginary axis},\ }\href {https://doi.org/10.1007/s10915-022-01963-z}
  {\bibfield  {journal} {\bibinfo  {journal} {J. Sci. Comput.}\ }\textbf
  {\bibinfo {volume} {92}},\ \bibinfo {pages} {107} (\bibinfo {year}
  {2022})}\BibitemShut {NoStop}%
\bibitem [{\citenamefont {Zhang}\ and\ \citenamefont
  {Gull}(2023)}]{zhang23arxiv}%
  \BibitemOpen
  \bibfield  {author} {\bibinfo {author} {\bibfnamefont {L.}~\bibnamefont
  {Zhang}}\ and\ \bibinfo {author} {\bibfnamefont {E.}~\bibnamefont {Gull}},\
  }\href {https://doi.org/10.48550/arXiv.2312.10576} {\bibinfo {title} {Minimal
  pole representation and controlled analytic continuation of {Matsubara}
  response functions}} (\bibinfo {year} {2023}),\ \Eprint
  {https://arxiv.org/abs/2312.10576} {arXiv:2312.10576 [cond-mat.str-el]}
  \BibitemShut {NoStop}%
\bibitem [{\citenamefont {Shinaoka}\ \emph {et~al.}(2020)\citenamefont
  {Shinaoka}, \citenamefont {Geffroy}, \citenamefont {Wallerberger},
  \citenamefont {Otsuki}, \citenamefont {Yoshimi}, \citenamefont {Gull},\ and\
  \citenamefont {Kune\v{s}}}]{Shinaoka20:tensornw}%
  \BibitemOpen
  \bibfield  {author} {\bibinfo {author} {\bibfnamefont {H.}~\bibnamefont
  {Shinaoka}}, \bibinfo {author} {\bibfnamefont {D.}~\bibnamefont {Geffroy}},
  \bibinfo {author} {\bibfnamefont {M.}~\bibnamefont {Wallerberger}}, \bibinfo
  {author} {\bibfnamefont {J.}~\bibnamefont {Otsuki}}, \bibinfo {author}
  {\bibfnamefont {K.}~\bibnamefont {Yoshimi}}, \bibinfo {author} {\bibfnamefont
  {E.}~\bibnamefont {Gull}},\ and\ \bibinfo {author} {\bibfnamefont
  {J.}~\bibnamefont {Kune\v{s}}},\ }\bibfield  {title} {\bibinfo {title}
  {Sparse sampling and tensor network representation of two-particle {Green's}
  functions},\ }\href {https://doi.org/10.21468/SciPostPhys.8.1.012} {\bibfield
   {journal} {\bibinfo  {journal} {SciPost Phys.}\ }\textbf {\bibinfo {volume}
  {8}},\ \bibinfo {pages} {12} (\bibinfo {year} {2020})}\BibitemShut {NoStop}%
\bibitem [{\citenamefont {Ge}\ \emph {et~al.}(2023)\citenamefont {Ge},
  \citenamefont {Halbinger}, \citenamefont {Lee}, \citenamefont {von Delft},\
  and\ \citenamefont {Kugler}}]{ge2023analytic}%
  \BibitemOpen
  \bibfield  {author} {\bibinfo {author} {\bibfnamefont {A.}~\bibnamefont
  {Ge}}, \bibinfo {author} {\bibfnamefont {J.}~\bibnamefont {Halbinger}},
  \bibinfo {author} {\bibfnamefont {S.-S.~B.}\ \bibnamefont {Lee}}, \bibinfo
  {author} {\bibfnamefont {J.}~\bibnamefont {von Delft}},\ and\ \bibinfo
  {author} {\bibfnamefont {F.~B.}\ \bibnamefont {Kugler}},\ }\href@noop {}
  {\bibinfo {title} {Analytic continuation of multipoint correlation
  functions}} (\bibinfo {year} {2023}),\ \Eprint
  {https://arxiv.org/abs/2311.11389} {arXiv:2311.11389 [cond-mat.str-el]}
  \BibitemShut {NoStop}%
\bibitem [{\citenamefont {Wallerberger}\ \emph {et~al.}(2023)\citenamefont
  {Wallerberger}, \citenamefont {Badr}, \citenamefont {Hoshino}, \citenamefont
  {Huber}, \citenamefont {Kakizawa}, \citenamefont {Koretsune}, \citenamefont
  {Nagai}, \citenamefont {Nogaki}, \citenamefont {Nomoto}, \citenamefont
  {Mori}, \citenamefont {Otsuki}, \citenamefont {Ozaki}, \citenamefont
  {Plaikner}, \citenamefont {Sakurai}, \citenamefont {Vogel}, \citenamefont
  {Witt}, \citenamefont {Yoshimi},\ and\ \citenamefont {Shinaoka}}]{SparseIR}%
  \BibitemOpen
  \bibfield  {author} {\bibinfo {author} {\bibfnamefont {M.}~\bibnamefont
  {Wallerberger}}, \bibinfo {author} {\bibfnamefont {S.}~\bibnamefont {Badr}},
  \bibinfo {author} {\bibfnamefont {S.}~\bibnamefont {Hoshino}}, \bibinfo
  {author} {\bibfnamefont {S.}~\bibnamefont {Huber}}, \bibinfo {author}
  {\bibfnamefont {F.}~\bibnamefont {Kakizawa}}, \bibinfo {author}
  {\bibfnamefont {T.}~\bibnamefont {Koretsune}}, \bibinfo {author}
  {\bibfnamefont {Y.}~\bibnamefont {Nagai}}, \bibinfo {author} {\bibfnamefont
  {K.}~\bibnamefont {Nogaki}}, \bibinfo {author} {\bibfnamefont
  {T.}~\bibnamefont {Nomoto}}, \bibinfo {author} {\bibfnamefont
  {H.}~\bibnamefont {Mori}}, \bibinfo {author} {\bibfnamefont {J.}~\bibnamefont
  {Otsuki}}, \bibinfo {author} {\bibfnamefont {S.}~\bibnamefont {Ozaki}},
  \bibinfo {author} {\bibfnamefont {T.}~\bibnamefont {Plaikner}}, \bibinfo
  {author} {\bibfnamefont {R.}~\bibnamefont {Sakurai}}, \bibinfo {author}
  {\bibfnamefont {C.}~\bibnamefont {Vogel}}, \bibinfo {author} {\bibfnamefont
  {N.}~\bibnamefont {Witt}}, \bibinfo {author} {\bibfnamefont {K.}~\bibnamefont
  {Yoshimi}},\ and\ \bibinfo {author} {\bibfnamefont {H.}~\bibnamefont
  {Shinaoka}},\ }\bibfield  {title} {\bibinfo {title} {sparse-ir: Optimal
  compression and sparse sampling of many-body propagators},\ }\href
  {https://doi.org/https://doi.org/10.1016/j.softx.2022.101266} {\bibfield
  {journal} {\bibinfo  {journal} {SoftwareX}\ }\textbf {\bibinfo {volume}
  {21}},\ \bibinfo {pages} {101266} (\bibinfo {year} {2023})}\BibitemShut
  {NoStop}%
\end{thebibliography}%

\end{document}